\documentclass[twocolumn,prd,superscriptaddress,nofootinbib]{revtex4-1}

\usepackage{amsfonts,amsmath,amssymb,mathrsfs}
\usepackage{hyperref}
\usepackage{color}
\usepackage{graphicx}  
\usepackage{dcolumn}   
\usepackage{bm}        
\usepackage[english]{babel}
\usepackage[bottom]{footmisc}
\usepackage{dblfnote}
\usepackage{mathtools}
\usepackage{amsmath}
\usepackage{physics}
\DFNalwaysdouble 

\def\a{\alpha}

\def\r{\rho}
\def\s{\sigma}
\def\t{\tau}
\def\m{\mu}
\def\n{\nu}
\def\k{\kappa}
\def\th{\theta}
\def\g{\gamma}\def\G{\Gamma}
\def\L{t}\def\l{V}
\def\D{\Delta}
\def\la{\langle}
\def\ra{\rangle}
\def\o{\omega}\def\O{\Omega}
\def\d{\delta}
\def\p{\partial}

\def\oxthree{{\cal O}(x^3) }

\def\half{\textstyle{\frac{1}{2}}}

\def\bdoc{\begin{document}}
\def\edoc{\end{document}}
\def\bea{\begin{equation}}
\def\eea{\end{equation}}

\def\beq{\begin{eqnarray}}
\def\eeq{\end{eqnarray}}
\def\be{\begin{eqnarray}}
\def\ee{\end{eqnarray}}
\def\ben{\begin{enumerate}}
\def\een{\end{enumerate}}
\def\la{\langle}
\def\ra{\rangle}
\def\a{\alpha}
\def\g{\gamma}\def\G{\Gamma}
\def\d{\delta}\def\D{\Delta}
\def\e{\epsilon}
\def\z{\zeta}

\def\th{\theta}
\def\k{\kappa}
\def\l{t}
\def\m{\mu}
\def\n{\nu}
\def\o{\omega}
\def\p{\pi}
\def\r{\rho}
\def\s{\sigma}
\def\t{\tau}
\def\L{{\cal L}}
\def\S{\Sigma }
\def\gsim{\; \raisebox{-.8ex}{$\stackrel{\textstyle >}{\sim}$}\;}
\def\lsim{\; \raisebox{-.8ex}{$\stackrel{\textstyle <}{\sim}$}\;}
\def\gtrsim{\gsim}
\def\lessim{\lsim}
\def\loc{{\rm local}}
\def\vm{v_{\rm max}}
\def\bh{\bar{h}}
\def\del{\partial}
\def\nab{\nabla}
\def\half{{\textstyle{\frac{1}{2}}}}
\def\fourth{{\textstyle{\frac{1}{4}}}}

\def\bD{{\bf D}}
\def\bE{{\bf E}}
\def\bF{{\bf F}}
\def\bB{{\bf B}}
\def\bP{{\bf P}}
\def\bV{{\bf v}}
\def\bv{{\bf v}}
\def\bx{{\bf x}}
\def\by{{\bf y}}
\def\bz{{\bf z}}
\def\ba{{\bf a}}
\def\bd{{\bf d}}
\def\bs{{\bf s}}
\def\bn{{\bf n}}
\def\bp{{\bf p}}

\def\O{\Omega}

\def\br{{\bf r}}
\def\bnab{{\bf \nab}}

\def\tE{\tilde{E}}
\def\tL{\tilde{L}}
\def\Horava{Ho\v{r}ava }

\def\oxtwo{\mathscr{O}\left(x^2\right)}
\def\oxthree{\mathscr{O}\left(x^3\right)}
\def\oxfour{\mathscr{O}\left(x^4\right)}
\def\oxfive{\mathscr{O}\left(x^5\right)}
\def\LL{\text{Lanczos-Lovelock}}

\def\ph{\phantom}

\begin{document}
\title{
Black Hole Zeroth Law in Higher Curvature Gravity}
\author{Rajes Ghosh}
\email{rajes.ghosh@iitgn.ac.in }
\affiliation{Indian Institute of Technology, Gandhinagar, Gujarat 382355, India.}
\author{Sudipta Sarkar}
\email{sudiptas@iitgn.ac.in}
\affiliation{Indian Institute of Technology, Gandhinagar, Gujarat 382355, India.}

\begin{abstract}
The zeroth law of black hole mechanics is an assertion of constancy of the surface gravity on a stationary Killing horizon. The Hawking temperature of the black hole horizon is proportional to the surface gravity. Therefore, the constancy of the surface gravity is reminiscent of the zeroth law of ordinary thermodynamics. In this work, we provide a proof of the zeroth law in Lanczos-Lovelock theories of gravity, where the Einstein Hilbert action is supplemented by higher curvature terms.

 \end{abstract}
\maketitle
\section{Introduction}
The laws of black hole mechanics have intriguing similarities with that of ordinary thermodynamics \cite{Bardeen}. The zeroth law in thermodynamics defines the concept of temperature via the transitivity of thermal equilibrium. Similarly, the zeroth law for black hole mechanics is an assertion of the constancy of surface gravity of a stationary Killing horizon. Once we associate the Hawking temperature of the black hole with the surface gravity, the zeroth law of black hole mechanics becomes the zeroth law of black hole thermodynamics. There are various possible proofs of the black hole zeroth law with different degree of generality. If we consider an event horizon in a stationary spacetime, the rigidity theorem \cite{Hawking:1971vc} assures that the event horizon is also a Killing horizon. Then, we can define the surface gravity using the time-like Killing vector which becomes null on the horizon. If this spacetime is a solution of Einstein's equation, the constancy of the surface gravity can be established using only the dominant energy condition on the matter field. No further assumption related to the geometric structure of the horizon is required.\\

However, quantum general relativity is a perturbatively non-renormalizable theory. Such a theory may make sense only as an effective theory with higher curvature terms supplementing the original action. It is also possible that the Einstein equations are modified at a shorter length scale, with new higher curvature terms. In principle, the nature of such terms will ultimately be decided by the details of the UV-physics. Nevertheless, we can always adopt a bottom-up approach and explore the aspects of some well-motivated higher curvature theories. A suitable example of such a theory is the $\LL$ gravity, where a specific form of the higher curvature terms is considered so that the field equations remain second order in time. Moreover, $\LL$ gravity is a unique ghost-free theory of gravity, where explicit black hole solutions are known \cite{Boulware:1985wk}. On the other hand, the derivation of the Hawking radiation is independent of the theory of gravity; it depends only on the geometric structure of the black hole spacetime \cite{Visser:2001kq}. Hence, it is important to understand the status of the zeroth law of black hole mechanics in higher curvature gravity, so that we can extend the black hole thermodynamics beyond general relativity. The aim of this work is to provide a proof of the zeroth law for stationary black hole solutions in $\LL$ gravity generalizing the result of general relativity. As per our knowledge, this is the first proof of black hole zeroth law for a well-motivated modified theory of gravity. This proof exemplifies the generality of the concepts of black hole mechanics beyond the Einstein's theory of gravity. \\

If we want to extend the proof of the zeroth law to higher curvature theories, we can not use the Einstein's field equations. The field equation will have contribution from higher curvature terms and further assumptions are used to establish the zeroth law. We can assume that the stationary space time contains a bifurcation surface; a compact cross section of the horizon in the past where the Killing field vanishes. Then, the zeroth law can be proven without the gravitational field equations \cite{Racz:1995nh}. Also, it turns out that the surface gravity is constant for all static black hole horizons \cite{Racz:1995nh}. Another possibility is to assume $t-\phi$ isometry; the existence of another space-like Killing vector which commutes with the original time-like one. The surface gravity is a constant for any black hole horizon if the spacetime has $t-\phi$ isometry \cite{Racz:1995nh, Waldb}. However, there is no general proof of the zeroth law once we venture beyond general relativity.\\

Interestingly, both the first law \cite{Wald, Iyer} and the quasi-stationary second law \cite{Chatterjee, Kolekar, Sarkar:2019xfd} have been extended for various other theories, like in $\LL$ gravity. At this stage, a natural question to ask is: whether the zeroth law can also be extended for general stationary black holes in $\LL$ class of theories. The importance of such an extension can hardly be overstated. This will establish that the thermodynamic properties of a stationary event horizon transcends general relativity and valid even for a higher curvature gravity. Note that, any such extension requires the validity of two separate theorems. First, the constancy of the surface gravity needs to be proven for a Killing horizon in the theory. Next, a proof of the rigidity theorem is also required to establish the zeroth law for the event horizon. In this work, we only consider the first problem. The validity of the rigidity theorem for $\LL$ gravity is still an important open issue.\\

Few years ago, an attempt was made \cite{Sarkar} to establish the zeroth law for $\LL$ gravity and the authors concluded that the constancy of surface gravity does not hold in general. In this work, we demonstrate that the aforesaid claim is not completely correct by proving the following statement: \textit{In $\LL$ theory, the surface gravity is a constant on a stationary Killing horizon, provided the matter obeys dominant energy condition.} The proof only requires an extra condition that the geometric quantities are well-behaved on the horizon, and there is smooth limit to general relativity. \\

The structure of the paper is as follows: we first review the properties of a Killing horizon. Next, the main result, i.e., the extension of the zeroth law to the Lanczos-Lovelock theory is presented. Finally, we conclude with further discussions. We work with the metric having mostly-plus or, $(-,+,+,+,....)$ signature and follow the same sign conventions as mentioned in \cite{Waldb}.

\section{Geometry of stationary event horizon}\label{redef}
Before going into the generalization of the zeroth law in higher curvature theories, we first make a quick review of the GR case. For this purpose, let's begin with the discussion of the properties of Killing horizons.\\ 
\\A Killing horizon is a null hypersurface $\mathcal{H}$ generated by a time-like Killing vector field $\xi^{a}=\left(\frac{\partial}{\partial v}\right)^{a}$, which is null on the horizon. Note that the parameter `v' along the generators is not necessarily an affine parameter. An important quantity related to a Killing horizon is the surface gravity ($\k$) which is defined by the equation at the horizon,
\bea \label{kappa}
\xi_{; a}^{b} \xi^{a}=\kappa\ \xi^{b}\ .
\eea
 By the virtue eq.(\ref{kappa}), it is possible to show that the surface gravity is constant along a generator \cite{Bardeen, Wald}, $\kappa_{; a}\ \xi^{a}=0$. However, surface gravity may vary from one generator to the other, in general. Therefore, it is interesting to study the variation of surface gravity across the generators. In order to proceed, we construct a basis $\{\xi^a, N^a, e^a_A \}$ on the Killing horizon. Here, $\xi^a$ is the Killing vector field discussed above and $N^a$ is another null vector satisfying, $\xi^a N_a = -1$; $ N^a N_a = 0$. 
Moreover, $\{ e^a_A \}$ are $(D-2)$ space-like vectors along the transverse directions satisfying, $e^b_A \xi_b = e^b_A N_b = 0.$ 
The Killing horizon of a stationary spacetime can be characterised by vanishing shear and expansion parameters. Therefore, using Raychaudhury's equation together with the evolution equation for shear, we obtain \cite{Waldb, Vega} on the horizon, $R_{a b} \xi^{a} \xi^{b}=0\ ,\, \, R_{a b c d}\ e_{A}^{a} e_{B}^{b} e_{C}^{c} \xi^{d}=0\ .$ With the help of these results, one can now compute the variation of surface gravity across the generators as,
\bea \label{across}
\kappa_{; a} e_{A}^{a}=-R_{a r p q}\ \xi^{r} N^{p} \xi^{q} e_{A}^{a}=-R_{a b}\ \xi^{a} e_{A}^{b}\ .
\eea

To establish the zeroth law, we need to show that the R.H.S of the Eq.(\ref{across} ) vanishes identically. For this purpose, we use the Einstein field equations in the context of GR and the dominant energy condition on the matter. These two conditions together imply that,
\bea \label{dom}
\kappa_{; a} e_{A}^{a}= 8 \pi j_{a} e^{a}_{A }\ ,\ \ \textrm{with}\ \ j_{a}=-T_{ab} \xi^b \propto \xi_a\ .
\eea
Then, the zeroth law follows directly from the orthogonality property, $e^b_A \xi_b =0$. It is evident that the proof is no longer valid when higher curvature terms modify the field equation. \\

\section{Aspects of Lanczos-Lovelock gravity}
The $\LL$ gravity is the most natural extension to the Einstein-Hilbert theory of gravity involving specific higher curvature terms in the action, so that the field equations remain second order in time. The Lagrangians of such theories can be compactly written as (neglecting the $m=0$ term, which is the cosmological constant) :
\bea \label{L}
\mathcal{L}^{(D)}=\sum_{m=1}^{[D-1) / 2]} \a_{m} \mathcal{L}_{m}^{(D)}\ ,
\eea
where $\{\a_m\}$ is a set of arbitrary constants with ${\a}_1=1$. The m-th order $\LL$ term, $\mathcal{L}_{m}^{(D)}$ is constructed from the D-dimensional curvature tensor $R^{cd}_{\ ab}$ and the generalized alternating tensor $\delta^{....}_{....}$ which is totally antisymmetric in both set of indices. The expression of $\mathcal{L}_{m}^{(D)}$ is given by 
\bea \label{Lm}
\mathcal{L}_{m}^{(D)}= \frac{1}{16 \pi} \frac{1}{2^{m}} \delta_{c_{1} d_{1} \ldots \epsilon_{m} d_{m}}^{a_{1} b_{1} \ldots a_{m} b_{m}} R_{a_{1} b_{1}}^{c_{1} d_{1}} \cdots R_{a_{m} b_{m}}^{c_{m} d_{m}}\ .
\eea
The terms correspond to m=1 and m=2 are the Einstein-Hilbert and Einstein-Gauss-Bonnet gravity (EGB) Lagrangians, respectively. The field equation of Lanczos-Lovelock theories can be expressed as, $G_{a b}+\sum_{j=2}^{m}\alpha_{j} E_{(j) a b}=8 \pi T_{a b}$, where
\bea \label{FE}
E_{(m) b}^{a}=-\frac{1}{2^{m+1}} \delta_{b c_1 d_{1} \ldots c_m d_m}^{a a_{1} b_{1} \ldots a_{m} b_{m}} R_{a_{1} b_{1}}^{c_{1} d_{1}} \cdots R_{a_{m} b_{m}}^{c_{m} d_{m}}\ .
\eea
More detailed discussion on $\LL$ gravity and the thermodynamical properties of corresponding black hole solutions can be found in \cite{Jacobson, Myers, Paddy}.

\section{Zeroth law in Lanczos-Lovelock gravity: A Previous Attempt}
In \cite{Sarkar}, the authors tried to extend the zeroth law of black hole thermodynamics for the $\LL$ theories of gravity. However, they claimed that such generalization does not work. We will show that their claim is not completely correct and one can indeed generalize the zeroth law in $\LL$ gravity.\\ 
\\In GR, we have easily shown that $\k_{;a} e^a_A = 0$ , using the Einstein’s field equation along with the dominant energy condition. However, for Lanczos-Lovelock gravity it is not so trivial to prove the same, even with the help of the field equation and the energy condition. The authors of \cite{Sarkar} formulated a method to tackle this issue. Proceeding in the same way as in GR (i.e., using the field equation and the dominant energy condition) , they finally arrived at the equation
\bea \label{acrossf}
\kappa_{; a} e_{B}^{a}= - \sum_{m>1}\ 2^m \alpha_{m}\ {}^{(D-2)} E_{(m-1) B}^{A}\ e_{A}^{a} R_{a r p q} \xi^{r} N^{p} \xi^{q}\ ,
\eea
\\where ${}^{(D-2)}E_{(m-1)B}^{A}$ is the equation of motion of the (m-1)-th order Lanczos-Lovelock theory constructed using intrinsic curvatures ${}^{(D-2)}R^{ABCD}$ of the horizon cross-section. Note that for GR (m=1), the right hand side of this equation vanishes trivially. However, for any other theory in $\LL$ class, the same can not be said. Thus, it is not at all obvious how the zeroth law can also be extended in $\LL$ theories.\\

Nevertheless, in the next section, we show that it is indeed possible to generalize the zeroth law for $\LL$ gravity. We only need a mild and reasonable assumption that there exists a well-defined limit to general relativity as $\alpha \to 0$. This will enable us to establish the zeroth law for stationary black holes in $\LL$ gravity without any further symmetry. We first work out the case of Einstein-Gauss-Bonnet gravity; the first non trivial $\LL$ term corresponding to the case $m = 2$.

\section{Generalization of the zeroth law in EGB theory ($m=2$)}
As a special case of eq.(\ref{acrossf}), we first consider $m=2$, i.e., the Einstein-Gauss-Bonnet (EGB) theory. Then, we have 
$${}^{(D-2)}E_{(m-1) B}^{A}= {}^{(D-2)}G_{B}^{A}= {}^{(D-2)}R_{B}^{A}-\frac{1}{2} \delta_{B}^{A}\ {}^{(D-2)}R \ .$$
Using eq.(\ref{across}), we can rewrite eq.(\ref{acrossf}) as follows (denoting $\alpha_{m=2}$ as $\a$, for brevity):
\begin{align} \label{m2}
\kappa_{; a} e_{B}^{a} = 4 \alpha\ {}^{(D-2)}G_{B}^{A}\left(\kappa_{; b} e_{A}^{b}\right)\ , \nonumber \\
\nonumber \\
\Rightarrow \left(\delta_{B}^{A}-4 \alpha\ {}^{(D-2)}G_{B}^{A}\right) \kappa_{; a} e_{A}^{a}=0\ .
\end{align}
The mathematical structure of this equation can be interpreted as follows: the vector $T_A = {\kappa}_{;a} e^a_A$ of the (D-2)-subspace, when acted upon by the square-matrix
 $M^{A}_{B}=\left( \delta^{A}_{B}-4 \alpha\ {}^{(D-2)}G^{A}_{B} \right),$ reduces to the zero-vector in that sub-space. If we demand that the EGB theory has a `smooth' limit (as $\a \to 0$) to GR, then one must have $M^{A}_{B} \to \delta^{A}_{B}\ .$ We want to emphasise that eq.(\ref{m2}) is an identity, and hence, true for all $\a$. In other words, if we vary $\a$, the value of the other parameters will change in such a manner that this equation still holds true.\\
\\One way to prove $\kappa_{;a} e^a_A=0$ will be to show
\bea \label{det}
\operatorname{det}\left(\delta_{B}^{A}-4 \alpha\ {}^{(D-2)}G_{B}^{A}\right) \neq 0
\eea
at every point on the hypersurface. However, it is a tremendous task to perform for an arbitrary choice of $\a$. For this reason, we first explore the case for small values of $\a$ and then use some other technique to prove the general case.\\
\\In the small $\a$ regime, we can expand the determinant as follows
\begin{align*}
\operatorname{det}\left(\delta_{B}^{A}-4 \alpha\ {}^{(D-2)} G_{B}^{A}\right) \approx1-4 \alpha \operatorname{tr}\left({}^{(D-2)}G_{B}^{A}\right) \\
=1+2 \alpha(D-4)\ {}^{(D-2)}R \ .
\end{align*}
At D = 5, the RHS of the above equation matches exactly with the entropy density \cite{Paddy} of the Gauss-Bonnet theory, which is surely non-zero. Thus, for 5-dimensional EGB theory, zeroth law holds true for small values of the coupling constant ($\a$) irrespective of its sign. On the other hand, at $D>5$, the positivity of entropy density implies the zeroth law for non-negative (small) values of $\a$ only. It is due to the following inequality satisfied by the determinant:  
$\operatorname{det}\left(\delta_{B}^{A}-4 \alpha\ {}^{(D-2)} G_{B}^{A}\right)> 2\a (D-5)\ {}^{(D-2)} R > 0$, at the first-order of $\a$.\\
\\Now, let's consider the case for arbitrary coupling constant $\a$ and general dimensions. To proceed, we rewrite eq.(\ref{m2}) in a very suggestive manner,
\bea \label{re}
4 \alpha\ {}^{(D-2)}G_{B}^{A}\ T_A = T_B\ .
\eea
The demand that in $\a \to 0$ limit we recover the correct GR result, puts a strong constraint on the plausible forms of $G^A_B$ and $T_A$ : These quantities must have `smooth' analytic structures in $\a$. That is, we should be able to write, $G^A_B= (G_0)^A_B+\a (G_1)^A_B+\a^2 (G_2)^A_B+....$ and similarly, $T_A= (T_0)_A + \a (T_1)_A + \a^2 (T_2)_A+....$, for all $\a$. The correct GR result ($T_A=0$) is obtained if we set $G^A_B \to (G_0)^A_B$ and $T_A \to (T_0)_A =0$, in $\a \to 0$ limit. It implies that there can not be any $\a^0$-order term in $T_A$, i.e., $(T_0)_A$ vanishes identically.\\
\\We can now substitute these forms into eq.(\ref{re}) for further analysis. Since eq.(\ref{re}) is true for all $\a$, we must compare the like-coefficients of various powers of $\a$ on the both sides of the equation. Then, it is easy to check that $(T_1)_A = (T_2)_A =....= 0$, i.e., the only solution is $T_A=0$.  This immediately leads to the zeroth law, $T_A=\kappa_{; a} e_{A}^{a}=0$ for the Einstein-Gauss Bonnet gravity.\\ 
\\Let's study a very suggestive illustration which seems to disagree with our previous conclusion in the first sight. Consider the case where $M^A_B=0$. In such situation, the equation $M^A_B T_A=0$ holds true even if $T_A \neq 0$. However, a closer inspection will show this is not really a disagreement. The point being the equation $M^A_B T_A=0$ is an identity and therefore, it is true even for GR. This observation, in turn, suggests that ${}^{(D-2)}G^A_B$ is regular in the limit $\a \to 0$, i.e., does not contain terms of order $\mathcal{O}(1/\a)$. With this constraint, it is almost trivial to see that $M^{A}_{B}=\left( \delta^{A}_{B}-4 \alpha\ {}^{(D-2)}G^{A}_{B} \right)$ can not always be zero (or, any fixed constant) at every point on the cross section as we vary $\a$. Thus, the only way $M^A_B T_A=0$ for every choice of $\a$ is $T_A=0$, which is in accordance with our previous result.

\section{Generalization of the zeroth law for $m>2$ Lanczos-Lovelock theories}
The Einstein-Gauss-Bonnet case, as discussed earlier, gives us the insight to extend the zeroth law even further for the full $\LL$ class of theories. The calculation in this section will be very similar to that in the previous section. For $m \geq 3$, the generalization of the of eq.(\ref{m2}) is given by,

\bea \label{mg2f}
\left(\delta_{B}^{A}- 2^{j} \alpha_j\ {}^{(D-2)}L_{B}^{A} \right) \kappa_{; a} e_{A}^{a}=0\ .
\eea
Here, we have defined a new quantity $L^{A}_{B}$ as follows
\bea \label{L}
L^{A}_{B}= \sum_{m \geq 2}\ 2^{m-j} \beta_{m,j}\ {}^{(D-2)} E_{(m-1) B}^{A}\ ,
\eea
where $\beta_{m,j}$ is a short-hand for $\left(\alpha_{m}/\alpha_{j} \right)$, with $\alpha_{j}$ being the first non-zero coupling constant in the set $\{\alpha_m \mid {m \geq 2}\}$. This equation has a very similar structure to that of eq.(\ref{m2}). Therefore, we can proceed in the same way as before and extend the zeroth law for the full $\LL$ class of theories.

\section{Conclusions}
Let us first summarize our result. We use the field equations of $\LL$ gravity and show that the surface gravity of a stationary Killing horizon is a constant, provided the matter obeys dominant energy condition and the higher curvature theory has a smooth limit to general relativity. The second condition seems reasonable and physical. However, this condition is non-trivial. This is because the space of solutions of any higher curvature gravity may be `larger' than that of general relativity. In fact, in EGB gravity, there are spherically symmetric solutions which do not have a smooth limit to general relativity \cite{Boulware:1985wk}. Our proof may not be valid for such pathological situations. \\

The proof clearly shows the special structure of the $\LL$ gravity, which allows us to write eq.(\ref{m2}) and (\ref{mg2f}), for the transverse derivatives of the surface gravity. Obviously, this structure may not be true for a general gravity theory. Nevertheless, we can still ask an interesting question: \textit{ What should be the structure of the field equations  of a general gravity theory, so that the zeroth law holds true as in the case of general relativity, without any further assumption including the assumption of the smooth limit}? To answer this question, we consider a theory of gravity which has an equation of motion, $G_{ab} + \alpha \, H_{ab} = 8\pi\,T_{ab}$, where $H_{ab}$ contains all the contributions from higher curvature terms. The GR limit is obtained as $\alpha \to 0$.  Now, extending the argument in section-(\ref{redef}), it is straightforward to show that the requirement of the validity of the zeroth law is $H_{ab} \xi^a e^{b}_{A}=0$ identically. Also, since we are working with a stationary Killing horizon, we have $ H_{ab} \xi^a \xi^b = 0$. This restricts the form of $H_{ab}$ on the horizon as:

\bea
H_{ab} = C\, \xi_a \xi_b +D\, g^{\perp}_{ab} + E_c N_{(a} \gamma_{b)}^{c} + F_{mn} \gamma^{m}_{a}\,\gamma^{n}_{b}\ .
\eea
Here, $\gamma^{a}_{b}$ is the intrinsic metric of the horizon cross section and $ g^{\perp}_{ab} $ is the metric on the $2$-plane perpendicular to it. Here, $C$, $D$, $E_c$ and $F_{mn}$ are local geometric quantities constructed from the metric and curvatures at the horizon. The validity of the zeroth law demands that the equation of motion of the higher curvature theory must be of this form on the horizon. Moreover, if we also have $t-\phi$ isometry, it will imply $H_{ab} \xi^a \propto \xi_b$. Then, we would have $E_c = 0$. Note that the $t-\phi$ isometry implies the zeroth law, but the opposite need not be true. It is remarkable that we can establish the zeroth law for the $\LL$ gravity, including general relativity without any such assumption of symmetry. This may not be the general feature of an arbitrary higher curvature gravity. Also, what remains to be proven is the analog of the \textit{Strong Rigidity Theorem} for $\LL$ gravity. The higher dimensional extension of Hawing's proof is provided in \cite{Hollands:2006rj}, where it is shown that in general relativity, a higher dimensional stationary black hole that is rotating must be axisymmetric. We expect that using our result, this proof may also be extended to the $\LL$ gravity. \\

It is intriguing that the structure of the field equations of $\LL$ gravity admits the generalization of the zeroth law. This result strengthen the idea that the thermodynamic nature of the black hole horizon is not limited to general relativity, but transcends to other theories of gravity. It would be interesting if there is a further generalization of this result to stationary black hole solutions of a general gravity theory.

\section*{Acknowledgement}
We want to thank Sumanta Chakraborty for extensive discussion and valuable comments on a previous draft. The research of SS is supported by the Department of Science and Technology, Government of India under the SERB Matrics Grant.
(MTR/2017/000399).


\begin{thebibliography}{100}



\section*{\bf{References}}    

\bibitem{Bardeen}
 J. M. Bardeen, B. Carter, S. W. Hawking, Commun. Math. Phys. 31, 161-170 (1973).

\bibitem{Hawking:1971vc}
S.~W.~Hawking,
Commun. Math. Phys. \textbf{25}, 152-166 (1972)
doi:10.1007/BF01877517


  
\bibitem{Boulware:1985wk}
D.~G.~Boulware and S.~Deser,
Phys. Rev. Lett. \textbf{55}, 2656 (1985)
doi:10.1103/PhysRevLett.55.2656

\bibitem{Visser:2001kq} 
  M.~Visser,
  Int.\ J.\ Mod.\ Phys.\ D {\bf 12}, 649 (2003)
  doi:10.1142/S0218271803003190
  [hep-th/0106111].
  
  
\bibitem{Racz:1995nh}
I.~Racz and R.~M.~Wald,
Class. Quant. Grav. \textbf{13}, 539-553 (1996)
doi:10.1088/0264-9381/13/3/017
[arXiv:gr-qc/9507055 [gr-qc]].


\bibitem{Waldb}
R. M. Wald, “General Relativity,” Chicago, Usa: Univ. Pr. ( 1984)

\bibitem{Wald}
R. M. Wald, Phys. Rev. D48, 3427-3431 (1993). [arXiv: gr-qc/9307038].


\bibitem{Iyer}
 V. Iyer and R. M. Wald, Phys. Rev. D 50, 846 (1994). [arXiv: gr-qc/9403028].


\bibitem{Chatterjee}
A. Chatterjee and S. Sarkar, Phys. Rev. Lett. 108, 091301 (2012) [arXiv:1111.3021 [gr-qc]].



\bibitem{Kolekar}
 S. Kolekar, T. Padmanabhan and S. Sarkar, arXiv:1201.2947 [gr-qc].

\bibitem{Sarkar:2019xfd}
S.~Sarkar,
Gen. Rel. Grav. \textbf{51}, no.5, 63 (2019)
doi:10.1007/s10714-019-2545-y
[arXiv:1905.04466 [hep-th]].


\bibitem{Paddy}
T. Padmanabhan and D. Kothawala,
Phys. Rept. \textbf{531}, 115-171 (2013)
doi:10.1016/j.physrep.2013.05.007
[arXiv:1302.2151 [gr-qc]].


\bibitem{Sarkar}
S. Sarkar and S. Bhattacharya,
Phys. Rev. D \textbf{87}, no.4, 044023 (2013)
doi:10.1103/PhysRevD.87.044023
[arXiv:1205.2042 [gr-qc]].


\bibitem{Ted}
T.~Jacobson, G.~Kang and R.~C.~Myers,
Phys. Rev. D \textbf{49}, 6587-6598 (1994)
doi:10.1103/PhysRevD.49.6587
[arXiv:gr-qc/9312023 [gr-qc]].


\bibitem{Weinberg}
S.~Weinberg, ``Gravitation and Cosmology,'' Wiley (1972).


\bibitem{Jacobson}
 T. Jacobson and R. C. Myers, Phys. Rev. Lett. 70, 3684
(1993) [arXiv:hep-th/9305016].


\bibitem{Myers}
R. C. Myers and J. Z. Simon, Phys. Rev. D 38, 2434
(1988).


\bibitem{Vega}
 I. Vega, E. Poisson and R. Massey, Class. Quant. Grav.
28, 175006 (2011) [arXiv:1106.0510 [gr-qc]].



\bibitem{Hollands:2006rj}
S.~Hollands, A.~Ishibashi and R.~M.~Wald,
Commun. Math. Phys. \textbf{271}, 699-722 (2007)
doi:10.1007/s00220-007-0216-4
[arXiv:gr-qc/0605106 [gr-qc]].









\end{thebibliography}
\end{document}